\documentclass[12pt, final,onecolumn] {IEEEtran}

\usepackage{epsfig,cite}
\usepackage{setspace}
\usepackage{subfigure}

\usepackage{graphicx}
\usepackage{longtable}
\usepackage{amsmath}
\usepackage{color}

\def\e{\begin{equation}}
\def\f{\end{equation}}

\title{An Efficient and Simple Analytical Model for Analysis of Propagation Properties in Impedance Waveguides}

\author{Olli Luukkonen, Constantin R.~Simovski,
Antti~V.~R\"ais\"anen, and Sergei A.~Tretyakov,
\thanks{
The work was supported in part by the Academy of Finland and Tekes
through the Center-of-Excellence Programme. Olli Luukkonen wishes to
thank the Nokia Foundation for financial support.}
\thanks{O.~Luukkonen, C.~R.~Simovski, A.~V.~R\"ais\"anen, and S.~A.~Tretyakov are with are with the Department of Radio Science and Engineering/SMARAD Center of Excellence, TKK Helsinki
University of Technology, P.O. 3000, FI-02015 TKK, Finland (email:
olli.luukkonen@tkk.fi)}}

\begin{document}

\maketitle {\center \large

}

\vspace{0.5cm}

\begin{abstract}

In this paper propagation properties of a parallel-plate waveguide
with tunable artificial impedance surfaces as sidewalls are studied
both analytically and numerically. The impedance surfaces comprise
an array of patches over a dielectric slab with embedded metallic
vias. The tunability of surfaces is achieved with varactors. Simple
design equations for tunable artificial impedance surfaces as well
as dispersion equations for the TE and TM modes are presented. The
propagation properties are studied in three different regimes: a
multi-mode waveguide, a single-mode waveguide, and below-cutoff
waveguide. The analytical results are verified with numerical
simulations.

\end{abstract}

\section{Introduction}

Artificial impedance surfaces
\cite{Kildal,Sievenpiper,Tretyakov_dynamicmodel,Tretyakov_mushroom,Simovski,Goussetis,Tretyakov_Analyticalmodelling,Luukkonen,Diaz}
have received a lot of interest since the beginning of the last
decade. In general the artificial impedance surfaces are composed of
a capacitive grid on top of a thin metal-backed dielectric
substrate. The substrate may include vias, as in
\cite{Tretyakov_dynamicmodel,Sievenpiper}, or may not include them
(\cite{Itoh,Maci,Goussetis}). Nevertheless, the purpose in both
types of designs is to use the grounded substrate to provide an
inductive response that together with the capacitive grid would
create a resonant structure. Because of the resonant nature such
impedance surfaces are commonly referred to as high-impedance
surfaces.

Recently some research has been devoted to electrical tunability of
high-impedance surfaces. The tunability has been realized e.g. by
connecting adjacent patches or strips to each other by voltage
controllable varactors in order to construct tunable antennas
\cite{Sievenpiper_beamsteering,Sievenpiper_leakywave}, phase
shifters \cite{Higgins}, lenses \cite{Xin}, band-pass filters
\cite{Higgins_bandpass}, and band-stop filters \cite{Xin_bandstop}.
In addition to the varactor-based tunable impedance surfaces also a
MEMS-based tunable high-impedance surface has been proposed in
\cite{Chicherin}. In this paper we concentrate on the varactor-based
tunable high-impedance surfaces.

In \cite{Sievenpiper_beamsteering,Xin,Sievenpiper_leakywave,
Higgins} the analysis of the tunable high-impedance surface has been
done by using a simplistic lumped-element model of the surface,
similar to that of \cite{Sievenpiper_thesis}. In addition to the
lumped element models, also a layered homogeneous material models
have been developed \cite{Clavijo,Hao} to predict the behavior of
non-tunable high-impedance surfaces. The lumped-element model
\cite{Sievenpiper_thesis} offers guidelines for the design of a
tunable high-impedance surface. For instance, in \cite{Mias} an
equivalent circuit model is used to approximate the effect of the
varactor resistance to the reflection characteristic of a
high-impedance surface similar to that described in
\cite{Sievenpiper_beamsteering}. However, the lumped-element model
cannot be used for an accurate analysis of the surface nor for the
analysis of many applications because it does not take the oblique
incidence into account.

Because of their unique characteristics the artificial impedance
surfaces have been used as wall coatings in different waveguiding
structures. For instance, close to the resonance frequency the input
surface impedance of an artificial impedance surface is high and the
surface behaves as a magnetic conductor. This feature has been
utilized in quasi-TEM waveguides \cite{Itoh,Higgins2,Lier,Skobelev}.
Furthermore, the possibility to electrically vary the input surface
impedance of an artificial impedance has been exploited in many
waveguiding applications
\cite{Higgins,Xin,Higgins_bandpass,Xin_bandstop}. For the design of
such applications accurate knowledge about the waveguide modes is
essential. In \cite{Kehn} analysis of dispersion in a rectangular
waveguide with impedance sidewalls comprising non-tunable
dipole-like frequency selective surfaces (FSS) on a metal-backed
dielectric slab has been done numerically by using the method of
moments.

In this paper we introduce a simple analytical model for a
varactor-tunable high-impedance surfaces that predicts the response
of the impedance surface very well even for oblique incidences. This
model is general and can be used for surfaces that comprise any type
of a rectangular patch array, for instance for such as those in
\cite{Sievenpiper_beamsteering,Sievenpiper_leakywave}. Together with
the dispersion equations, the model for the tunable impedance
surfaces is used to study the propagation properties of a
parallel-plate waveguide having either one or two tunable impedance
surfaces replacing the metallic plates. In particular,
parallel-plate waveguides allowing multi-mode or single-mode
propagation are studied. Also a waveguide operating below its cutoff
frequency is considered. Numerical full-wave simulations verify the
analytical results and show that the used analytical model describes
accurately the properties of tunable impedance surfaces in
waveguiding set-ups.

\section{Analytical model}

\begin{figure}[b]
\centering \includegraphics[width=7cm]{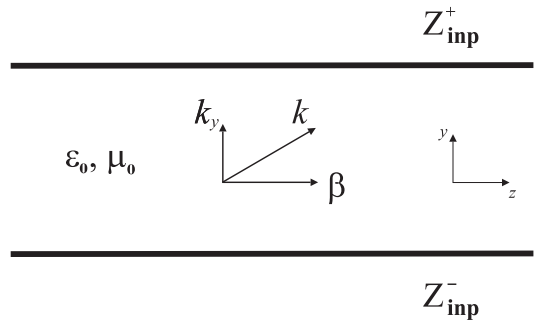}
\caption{Illustration of the two dimensional waveguide confined by
two impedance surfaces.} \label{fig:2dwaveguide}
\end{figure}

We study the propagation properties of a parallel-plate waveguide
having tunable impedance surfaces using the plane-wave
interpretation. The two-dimensional waveguide geometry is
illustrated in Fig.~\ref{fig:2dwaveguide}. The number of the
transversal wave numbers is limited to one as no propagation takes
place in the $x$-direction. The waveguide is confined in the
$y$-direction by plates that can be modeled with an impedance
surface $Z_{\rm inp}^+$ or $Z_{\rm inp}^-$ that are dependent both
on the wave number $k$ and the propagation constant along the
waveguide $\beta$. The notation $\pm$ refers to the upper/lower
surface, respectively.

The dispersion relations for the parallel-plate waveguide presented
in Fig.~\ref{fig:2dwaveguide} can be solved from the boundary
conditions at the waveguide sidewalls and from the transverse
components of the magnetic and electric fields. The transverse
components can be calculated from the longitudinal ($z$-) components
using general plane-wave solutions in the $y$-dimension (see e.g.
\cite{Pozar}). The resulting dispersion equation for the TE modes
can be written as: \e \tan(k_{\rm y}d) = j\eta\frac{k}{k_{\rm
y}}\frac{Z_{\rm inp}^+ + Z_{\rm inp}^-}{\eta^2\frac{k^2}{k_{\rm
y}^2} + Z_{\rm inp}^+Z_{\rm inp}^-}. \label{eq:dispTElast} \f

For TM modes the dispersion equation reads \e \tan(k_{\rm y}d) =
j\eta\frac{k_{\rm y}}{k}\frac{Z_{\rm inp}^+ + Z_{\rm
inp}^-}{\eta^2\frac{k_{\rm y}^2}{k^2} + Z_{\rm inp}^+Z_{\rm inp}^-}.
\label{eq:dispTMlast}\f In the above formulae $k_{\rm
y}=\pm\sqrt{k^2-\beta^2}$ is the transverse wave number, and $\eta$
is the plane-wave impedance of the medium filling the waveguide. By
choosing the ``minus''-branch of the transverse wave number, also
surface modes on the waveguide sidewalls are predicted by the above
dispersion equations.

\subsection{The input surface impedance}

\begin{figure}[t!]
\centering
\subfigure[]{\includegraphics[width=7.5cm]{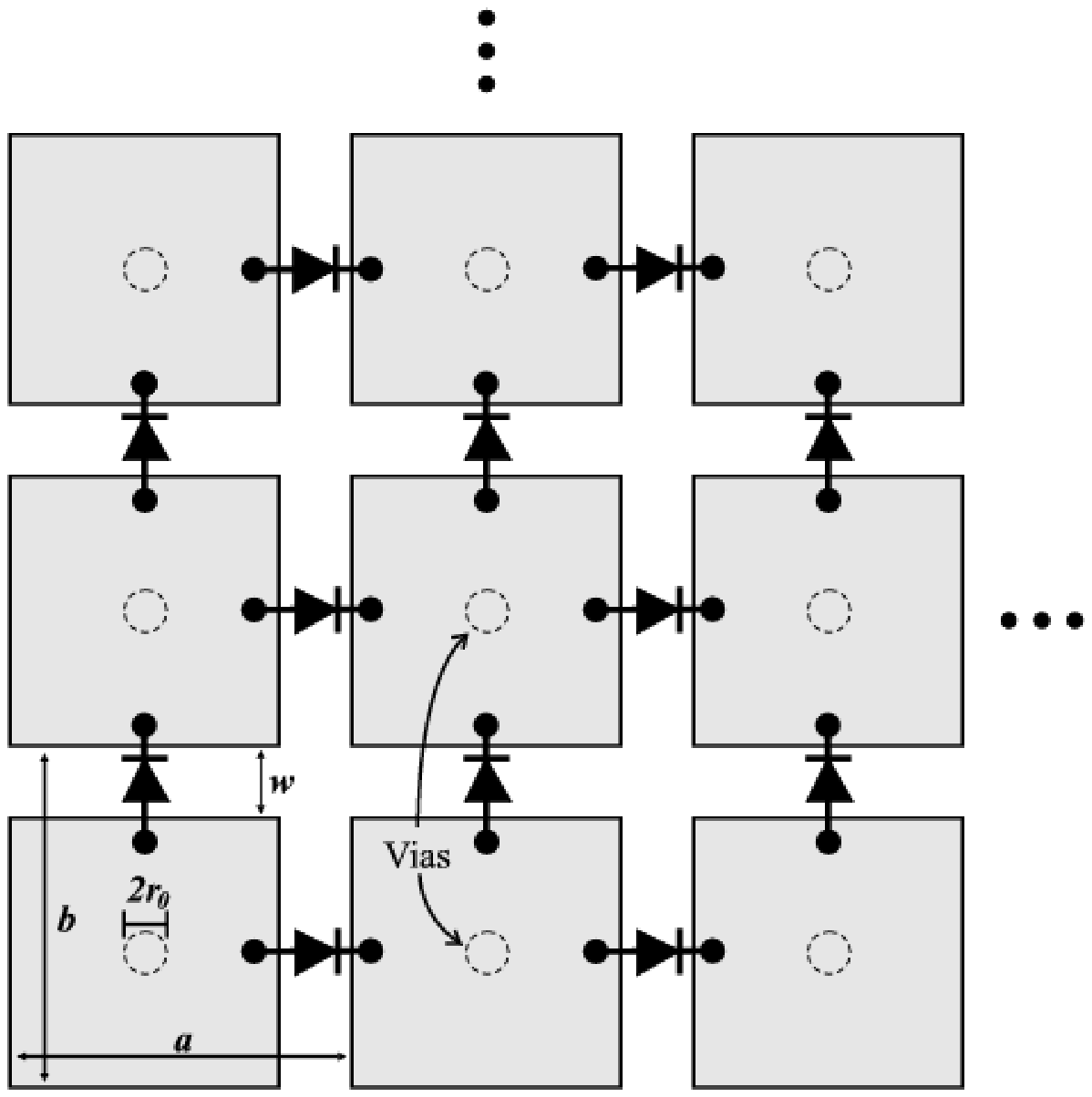} }
\subfigure[]{\includegraphics[width=7.5cm]{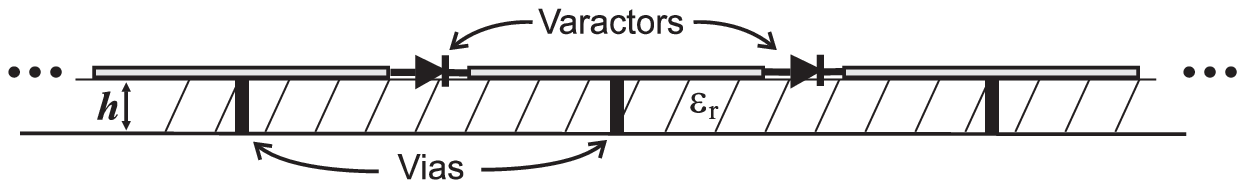}}
\caption{(a) A view from the top of the Sievenpiper mushroom
structure loaded with varactors. (b) A view from the side.}
\label{fig:mushroom_structure}
\end{figure}

\begin{figure}[t!]
\centering \includegraphics[width=7.5cm]{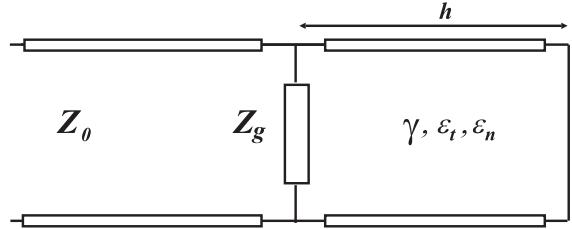}
\caption{A transmission-line model for modeling artificial impedance
surfaces. $Z_0$ is the free-space impedance.}
\label{fig:transmissionlinemodel}
\end{figure}

The dispersion equations \eqref{eq:dispTElast} and
\eqref{eq:dispTMlast} have been derived for arbitrary surface
impedances. In this paper the tunable version of a mushroom-type
artificial impedance surface (see
Fig.~\ref{fig:mushroom_structure}), proposed in \cite{Sievenpiper},
is studied as a possible particular realization. The metallic plates
of a waveguide are replaced with artificial impedance surfaces that
are comprised of metallic rectangular patches and metal-backed
dielectric substrates with embedded vias. The tunability is achieved
by connecting the adjacent rectangular patches to each other by
varactors. The input surface impedance of an artificial impedance
surface can be modeled through a transmission-line model shown in
Fig.~\ref{fig:transmissionlinemodel}. The input impedance is hence a
parallel connection of the grid impedance of an array of rectangular
patches and the surface impedance of a metal-backed dielectric slab
with embedded vias: \e Z_{\rm inp}^{-1} = Z_{\rm g}^{-1} + Z_{\rm
s}^{-1}.\f In the above equation subscript ${\rm g}$ refers to the
grid impedance of an array of rectangular patches and ${\rm s}$
refers to the surface impedance of the substrate.

A simple and accurate analytical model for a non-tunable
mushroom-type impedance surface is available \cite{Luukkonen}. In
\cite{Luukkonen} the mushroom structures comprised arrays of patches
on top of a dielectric layer. However, in this paper we consider
mushroom structures that comprise arrays of patches on top of a
metal-backed dielectric slab with embedded metallic vias. The vias
are needed to provide the bias voltage for the varactors that are
used to vary the capacitance between the adjacent strips or patches
(as in \cite{Higgins,Sievenpiper_beamsteering}). For TE modes the
electric field component is perpendicular to the vias and, in the
case of thin vias, does not excite them. It can be concluded that
the analytical model \cite{Luukkonen} can be readily applied for TE
modes in the case of embedded vias by taking the effect of varactors
into account correctly. However, for TM modes the electric field has
a parallel component to the vias. In this case it is possible to
model the metallic vias in a dielectric slab as an effective wire
medium \cite{Diaz,Tretyakov_Analyticalmodelling}.

\subsection{Surface impedance of the dielectric slab with embedded
vias}

The surface impedance for a wire medium comprising thin perfectly
conducting wires reads \cite{Tretyakov_Analyticalmodelling}: \e
Z_{\rm s}^{\rm TM} = j\omega\mu_0\frac{\tan(\gamma_{\rm
TM}h)}{\gamma_{\rm TM}}\frac{k^2 - \beta^2 - k_{\rm p}^2}{k^2 -
k_{\rm p}^2}, \f where $k=k_0\sqrt{\varepsilon_{\rm r}}$ is the wave
number in the host medium, \e \gamma_{\rm TM}^2 =
\omega^2\varepsilon_0\varepsilon_{\rm t}\mu_0 -
\frac{\varepsilon_{\rm t}}{\varepsilon_{\rm n}}\beta^2,\f
$\varepsilon_{\rm t}$ is the relative permittivity for the fields
along the transverse plane, and \e k_{\rm p} =
\frac{1}{a\sqrt{\frac{1}{2\pi}\ln{\frac{a^2}{4r_0(a - r_0)}}}}. \f
Furthermore, $a$ is the period of the wires, $r_0$ is the radius of
the wires, and the relative permittivity for the fields along the
normal of the medium reads \e \varepsilon_{\rm n} = \varepsilon_{\rm
t}\left(1 - \frac{k_{\rm p}^2}{k^2\varepsilon_{\rm t}}\right). \f In
the case when the vias are thin and vertically oriented, the
relative permittivity for the fields along the transversal plane,
$\varepsilon_{\rm t}$, equals to the relative permittivity of the
host medium, $\varepsilon_{\rm r}$.

For TE modes the electric field is perpendicular to the thin
metallic wires. In this case, as discussed above, the electric field
does not excite the wires, and the surface impedance for the TE mode
is that of a metal-backed dielectric slab (see, e.g.,
\cite{Tretyakov_Analyticalmodelling}): \e Z_{\rm s}^{\rm TE} =
j\omega\mu_0\frac{\tan(k_{\rm y}\sqrt{\varepsilon_{\rm r}}h)}{k_{\rm
y}\sqrt{\varepsilon_{\rm r}}}. \f

\subsection{The grid impedance}

The grid impedance for an array of patches can be calculated through
the approximative Babinet principle using the averaged boundary
conditions for a mesh of wires or strips. The averaged boundary
condition for a mesh of strips is available e.g., in
\cite{Tretyakov_Analyticalmodelling}. The grid impedance for an
array of ideally conducting patches on top of a dielectric substrate
reads \cite{Luukkonen}: \e Z_{\rm g}^{\rm TM} = -j\frac{\eta_{\rm
eff}}{2\alpha}, \label{eq:array_impedance_TM}\f \e Z_{\rm g}^{\rm
TE} = -j\frac{\eta_{\rm eff}}{2\alpha\left(1 - \frac{k_0^2}{k_{\rm
eff}^2}\frac{\sin^2\theta}{1 + \frac{b}{a}}\frac{b}{a} \right)},
\label{eq:array_impedance_TE}\f where the effective wave impedance
$\eta_{\rm eff} = \frac{\eta_0}{\sqrt{\varepsilon_{\rm eff}}}$, the
effective wave number  $k_{\rm eff} = k_0\sqrt{\varepsilon_{\rm
eff}}$, $\sin^2(\theta) = \frac{k_0^2 - k_{\rm y}^2}{k_0^2}$, and
$b$ and $a$ are the dimensions of the unit cell of the structure
along $x-$ and $z-$axis, respectively. Further, $\alpha$ is the {\it
grid parameter}: \e \alpha = \frac{k_{\rm
eff}b}{\pi}\ln\left(\frac{1}{\sin\left(\frac{\pi w}{2b} \right)}
\right), \label{eq:alpha}\f where $w$ is the gap between the
adjacent patches (see Fig.~\ref{fig:mushroom_structure}). The
limitations for above expressions of the grid impedance have been
discussed in more detail in \cite{Luukkonen}. It can be concluded
here that \eqref{eq:array_impedance_TM} and
\eqref{eq:array_impedance_TE} are valid when $w<a,b$ and up to the
frequencies when $a,b \approx \frac{\lambda}{2}$. The effective
relative permittivity for the array of patches or grid of strips on
the boundary between two medium having relative permittivities of
$\varepsilon_1$ and $\varepsilon_2$ reads approximately \cite
{Compton_strips}: \e \varepsilon_{\rm eff} = \frac{\varepsilon_1 +
\varepsilon_2}{2}. \label{eq:effective_permittivity}\f

In this paper the array of patches is located on a boundary between
free space and wire medium. From above it is known that in a wire
medium the fields along the transversal and normal axis see
different effective relative permittivities. For an array of patches
the electric fields are concentrated mainly between the adjacent
patches, transverse with respect to the vias, and the effect of the
vias on the electric response is weak. For this reason the
transversal relative permittivity of the wire medium is used in
\eqref{eq:effective_permittivity}.

The grid impedances \eqref{eq:array_impedance_TM} and
\eqref{eq:array_impedance_TE} can be written in a lumped-element
form as: \begin{equation} Z_{\rm g}^{\rm TM, TE} = \frac{1}{j\omega
C_{\rm g}^{\rm TM, TE}}, \label{eq:lumped_impedance} \end{equation}
where $C_{\rm g}^{\rm TM, TE}$ is the grid capacitance for the TM-
or TE-polarized incidence fields. The grid capacitance equals to the
averaged capacitance per one unit cell in the $x-$ and $z-$direction
for the TM- and TE-polarized cases. Using
\eqref{eq:array_impedance_TM}, \eqref{eq:array_impedance_TE}, and
\eqref{eq:lumped_impedance} the grid capacitance for an array of
patches can be written as:
\begin{equation} C_{\rm g}^{\rm TM} =
\frac{b\varepsilon_0\left(\varepsilon_1 + \varepsilon_2
\right)}{\pi}\ln\left(\frac{1}{\sin\left(\frac{\pi
w}{2b}\right)}\right) \label{eq:array_capacitance_TM},
\end{equation}
\begin{equation} C_{\rm g}^{\rm TE} =
\frac{b\varepsilon_0\left(\varepsilon_1 +
\varepsilon_2\right)}{\pi}\ln\left(\frac{1}{\sin\left(\frac{\pi
w}{2b}\right)}\right)\left(1 - \frac{k_0^2}{k_{\rm
eff}^2}\frac{\sin^2\theta}{1 + \frac{b}{a}}\frac{b}{a} \right)
\label{eq:array_capacitance_TE}. \end{equation}

The above formulas for the capacitive grid impedances hold for
ideally conducting patches. Although the capacitive impedance is
derived through averaged fields on the grid, we may consider a
lumped-element model here. This way the capacitance of the
varactors, that are used for tuning, can be included in the analysis
easily: The additional capacitance of the varactors is connected in
parallel with the grid capacitance and the total impedance of a unit
cell is thus a parallel connection of these impedances. Hence, the
total grid capacitance of the array of patches with varactors can be
written as:
\begin{equation} C_{\rm tot}^{\rm TM, TE} = C_{\rm g}^{\rm TM, TE} + C_{\rm var}, \label{eq:total_capacitance}
\end{equation} where $C_{\rm g}$ is the grid capacitance for an array of ideally
conducting patches and $C_{\rm var}$ is the capacitance of the
tunable varactor. Finally, the total grid impedance reads:
\begin{equation} Z_{\rm g}^{\rm TM, TE} = \frac{1}{j\omega C_{\rm tot}^{\rm TM, TE}} =
\frac{1}{j\omega\left(C_{\rm g}^{\rm TM, TE} + C_{\rm var}\right)}.
\end{equation}

In order to most effectively tune the grid impedance, the
capacitance of the varactor needs obviously to be considerably
larger than the grid capacitance.

The effect of the varactor resistance to the performance of a
tunable high-impedance surface has been studied in \cite{Mias}. It
is possible to include the effect of the intrinsic diode resistance
to the above analysis in a similar way as done in \cite{Mias}.
However, this is considered to be out of the scope of this paper.

\begin{figure}[t!]
\centering
\includegraphics[width=0.3\textwidth]{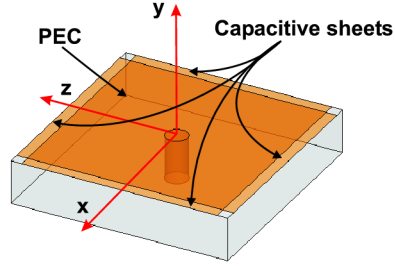}
\caption{Color online. Simulation model of the tunable
high-impedance surface unit cell. Because of the periodicity, the
capacitance of each capacitive sheet equals to $2C_{\rm var}$.}
\label{fig:simulation model}
\end{figure}

\begin{figure}[t!]
\centering
\includegraphics[width=0.5\textwidth]{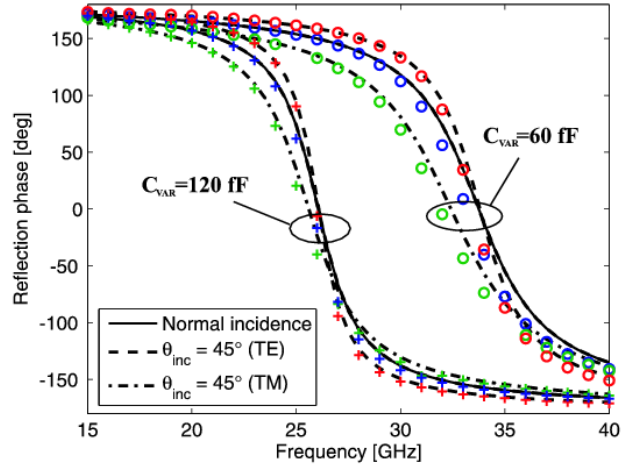}
\caption{Color online. The reflection phases for different
incidences and for different values of varactor capacitance. HFSS
simulation results are denoted with crosses and circles for varactor
capacitance values of 60\,fF and 120\,fF, respectively. The
simulation results for normal incidence, $45^\circ$ (TM), and
$45^\circ$ (TE) are colored blue, green, and red, respectively.}
\label{fig:reflectionphase}
\end{figure}

\section{Numerical validation and results}

In this section the propagation properties of a parallel-plate
waveguide having either one or two artificial impedance surfaces are
studied analytically. The analytical results are verified with
simulations using Ansoft's High Frequency Structure Simulator
(HFSS).

The impedance surfaces for the waveguides are designed for the lower
millimeter wave region, namely Ka-band (26--40~GHz). Following the
notations in Fig.~\ref{fig:mushroom_structure}, the parameters of
the studied artificial impedance surface are: $a=b=1$\,mm,
$w=0.1$\,mm, $h=0.2$\,mm, and $\varepsilon_{\rm r} = \varepsilon_2 =
4$. The medium inside the impedance waveguide is air ($\varepsilon_1
= 1$). As in any resonant circuit, the bandwidth of the resonance
can be increased by increasing the effective inductance (the height
of the grounded dielectric slab) with respect to the effective
capacitance. Here the resonance bandwidth of the high-impedance
surface is decreased intentionally so that the effects due to the
resonance are clearly recognizable in the dispersion plots. The
varactors on each edge of the patch are modeled with lumped
capacitive sheets whose value of capacitance is changed depending on
the studied case. The simulation model of the high-impedance surface
is shown in Fig.~\ref{fig:simulation model}. The periodicity in the
simulation model is achieved by using the periodical boundary
conditions available in HFSS.

\begin{figure}[t!]
\centering
\subfigure[]{\includegraphics[width=0.5\textwidth]{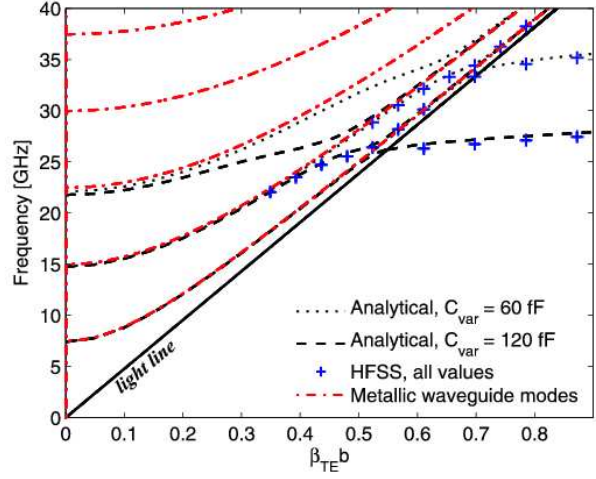}
}
\subfigure[]{\includegraphics[width=0.5\textwidth]{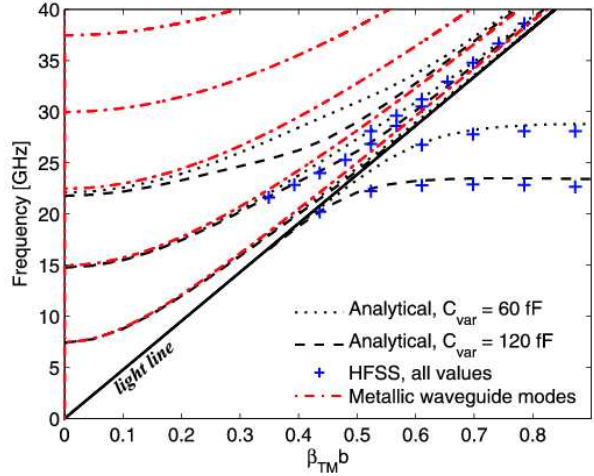}}
\caption{Color online. (a) The propagation properties for TE modes
in an impedance waveguide with one tunable impedance surface. (b)
The propagation properties for TM modes in an impedance waveguide
with one tunable impedance surface. The fundamental modes of metal
waveguide are plotted with dash-dot lines. $\beta_{\rm TE}$ and
$\beta_{\rm TM}$ refer here to the propagation constants of the TE
and TM modes, respectively. Only the two lowest modes are simulated.
} \label{fig:dispersion1}
\end{figure}

According to \eqref{eq:array_capacitance_TM} the grid capacitance of
the designed impedance surface is approximately 26\,fF. Based on
this information and knowing the frequency band of interest, the
capacitance of the varactors is varied from 60\,fF to 120\,fF; with
these values the resonance frequency of the surface for the normal
incidence appears to be approximately at 34\,GHz and 26\,GHz,
respectively. Furthermore, the capacitance of the varactors is
considerably larger than the grid capacitance, as discussed earlier.

The reflection phases of the surface for different values of $C_{\rm
var}$ and for different incident angles are shown in
Fig.~\ref{fig:reflectionphase}. The analytical results concur with
the simulation results very accurately. Clearly the bandwidth for
the TE polarization becomes smaller and the bandwidth for the TM
polarization larger as the angle of incidence grows. The effect of
this to the propagation properties of the impedance waveguide will
be discussed later.

\subsection{A multi-mode waveguide}

Before moving on to the interpretation of the dispersion plots, the
terms used in this and following paragraphs need to be clarified. It
is natural to use the modes propagating in an empty metallic
parallel-plate waveguide as reference cases when studying the
propagation properties of an impedance waveguide. Therefore the
modes propagating in a metallic waveguide will be referred as {\it
fundamental modes} from here on in order to distinguish them from
the modes of an impedance waveguide (referred plainly as {\it
modes}).

The dispersion curves of a 20-mm high parallel-plate waveguide with
one tunable impedance surface are shown in
Fig.~\ref{fig:dispersion1}. The second surface is perfectly
conducting metal. The height of the waveguide was chosen so that
many fundamental modes would propagate in the waveguide at Ka-band.
The fundamental modes of a parallel-plate waveguide are shown with
dash-dot lines. The simulation results have been marked by crosses
in Fig.~\ref{fig:dispersion1}. The concurrence between the
analytical and numerical results is very good.

\begin{figure}[t!]
\centering \mbox{
\subfigure[]{\includegraphics[width=0.15\textwidth,height=5cm]{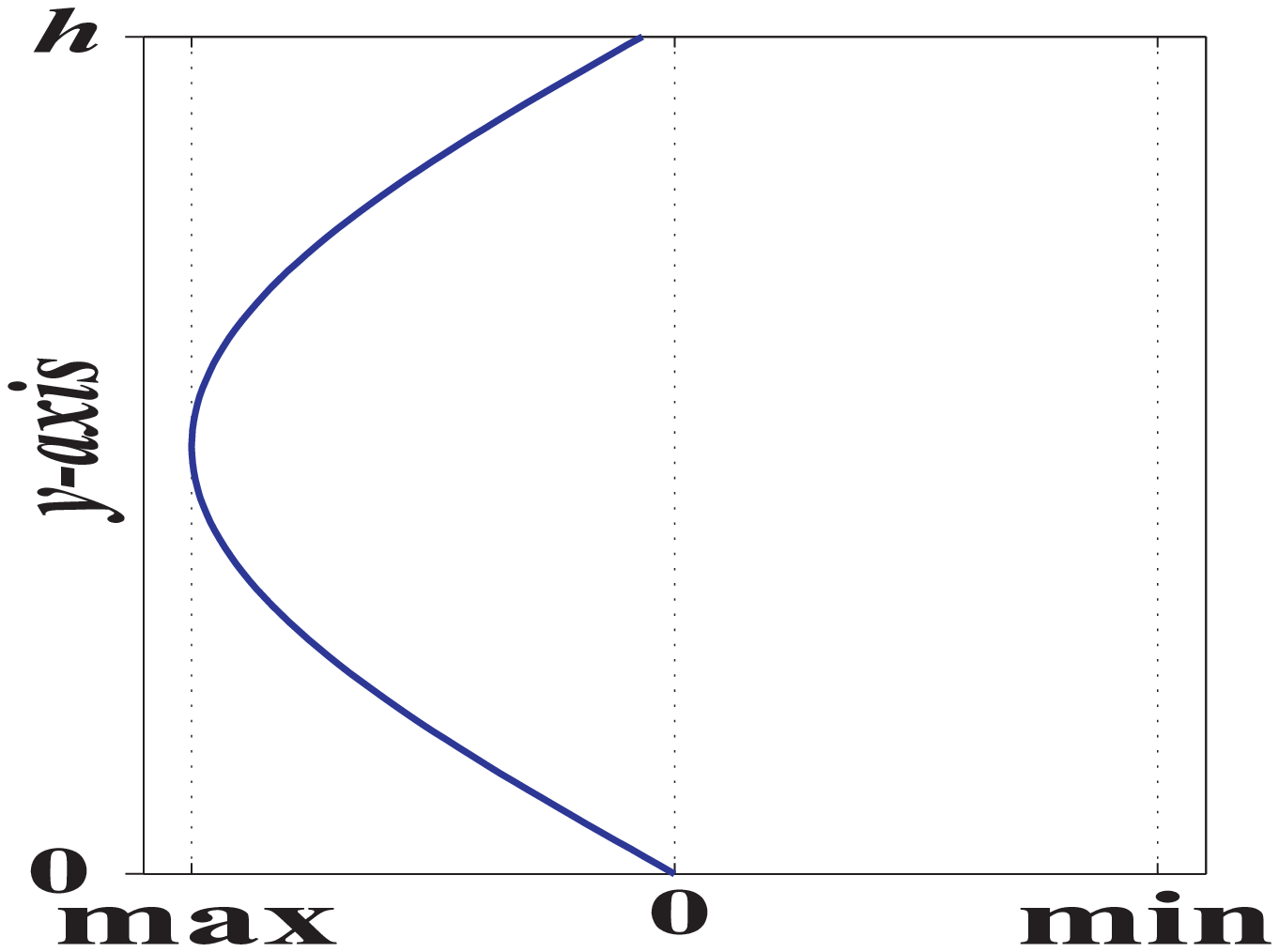}
}
\subfigure[]{\includegraphics[width=0.15\textwidth,height=5cm]{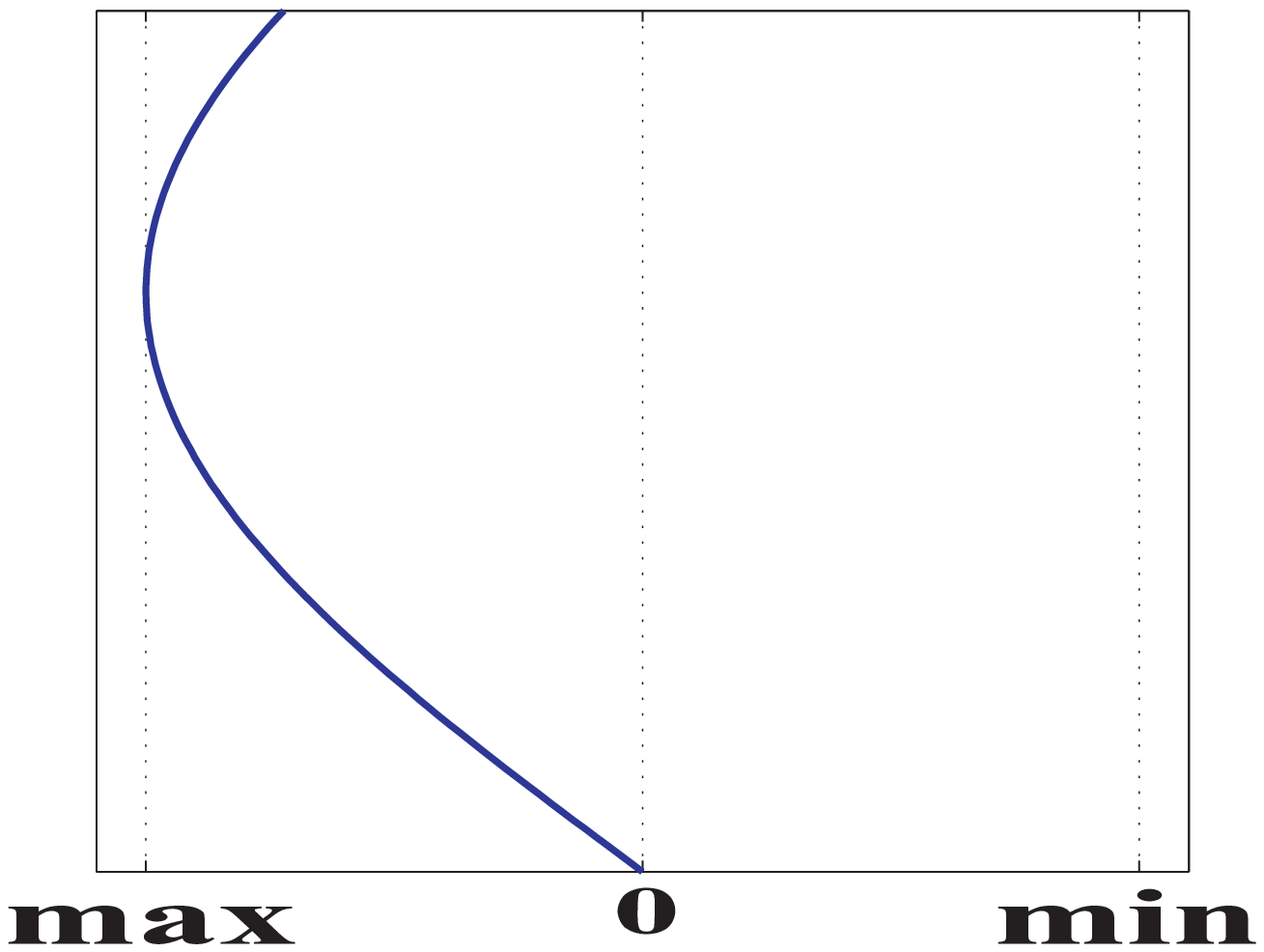}}
\subfigure[]{\includegraphics[width=0.15\textwidth,height=5cm]{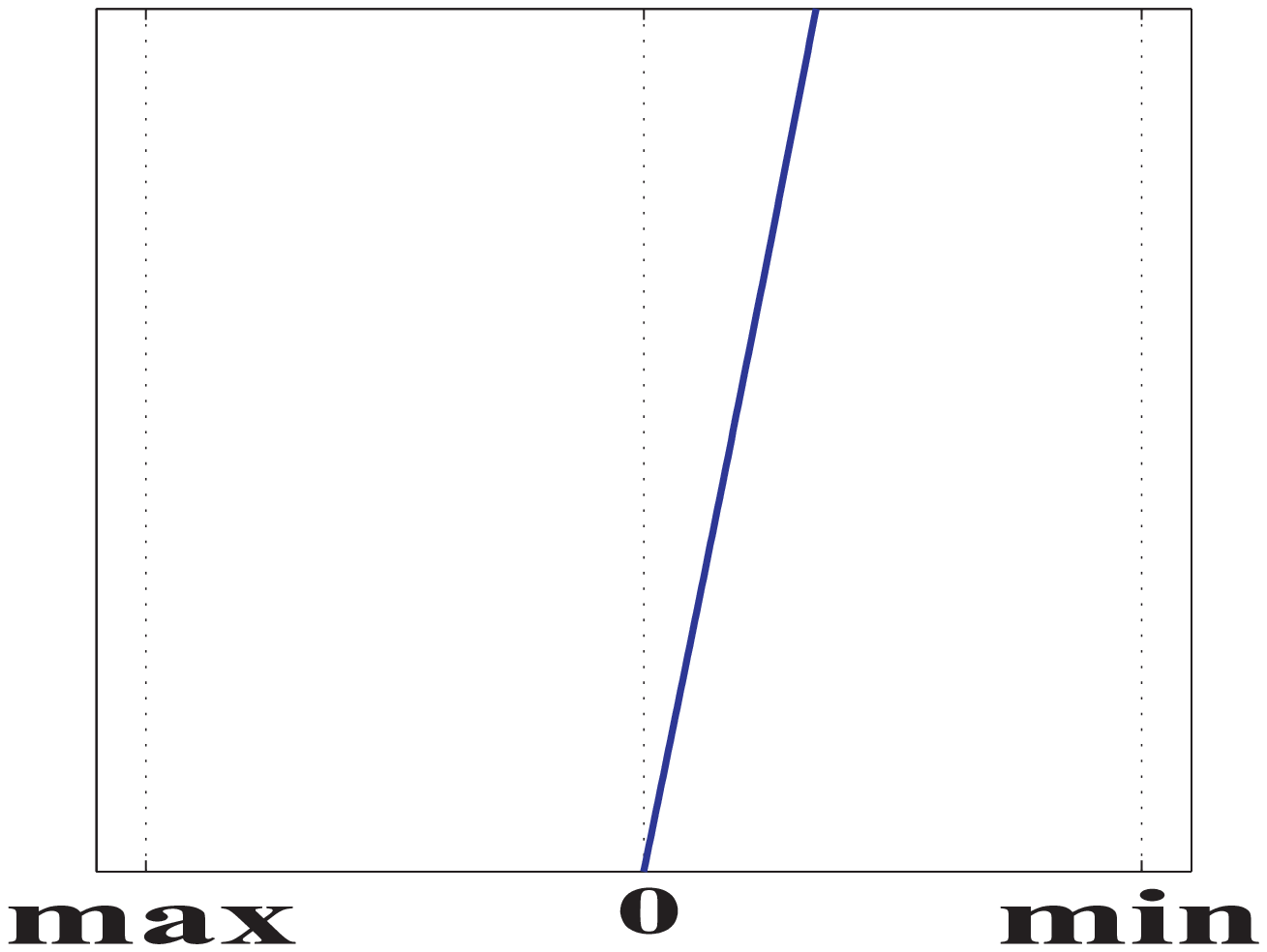}}}
\caption{The normalized magnitude of the $x-$component of the
electric field for the ${\rm TE}_{\rm 01}$ mode at points (a) $\beta
b=0.5$, $f = 25.0$\,GHz (b) $\beta b=0.69$, $f = 33.5$\,GHz, and (c)
$\beta b=0.71$, $f = 34.0$\,GHz  } \label{fig:fieldpattern}
\end{figure}

In Fig.~\ref{fig:dispersion1} near the resonance frequency of the
impedance surface a $180^\circ$ mode hop is observed for both
polarizations. The propagating wave exhibits a $180^\circ$ phase
shift while the mode morphs from one fundamental mode to another.
This is due to the change of the reflection phase of the impedance
surface. For instance, a wave propagating at $27.5$\,GHz near the
second fundamental mode, ${\rm TE}_{02}$ in
Fig.~\ref{fig:dispersion1}(a), morphs into the first fundamental
mode, ${\rm TE}_{01}$, when the varactor capacitance is changed
gradually from 60\,fF to 120\,fF over a certain distance. This is
similar to the behavior of corrugated-waveguide mode converters,
where the depth of the corrugations is tapered gradually
\cite{Thumm}. Similar behavior occurs for higher-order modes and for
${\rm TM}_{\rm 01}$ as well. The field pattern of the ${\rm TE}_{\rm
01}$ mode is plotted in Fig.~\ref{fig:fieldpattern}.

The mode conversion is seen to happen more gradually for the TM
modes in Fig.~\ref{fig:dispersion1} (b) than for the TE modes in
Fig.~\ref{fig:dispersion1} (a). This is because the bandwidth of the
high-impedance surface becomes wider for TM polarized fields than
for TE polarized fields as the angle of incidence grows, as
discussed above. This creates an advantage for the TE modes over the
TM modes in tunable impedance waveguide applications: The needed
range of tuning of the resonant frequency of the high-impedance
surface is smaller for the TE modes than for the TM modes in
impedance waveguide applications.

\begin{figure}[t!]
\centering
\subfigure[]{\includegraphics[width=0.5\textwidth]{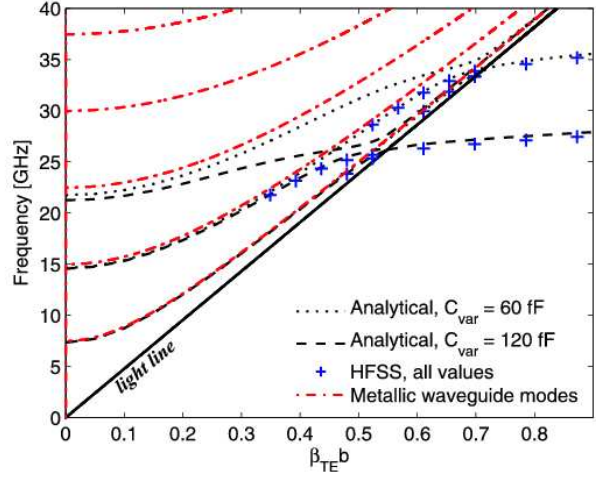}
}
\subfigure[]{\includegraphics[width=0.5\textwidth]{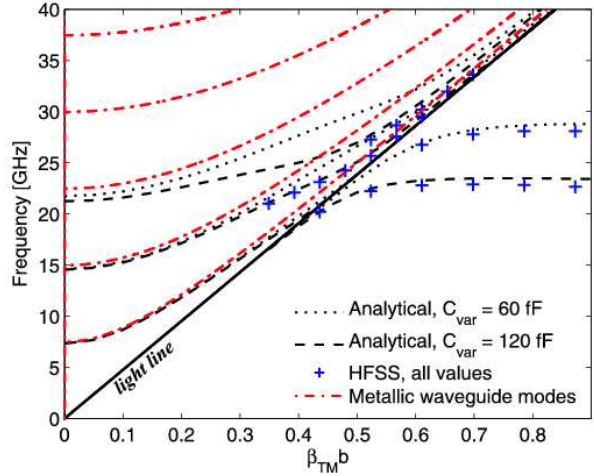}}
\caption{Color online. (a) The propagation properties for TE modes
in an impedance waveguide with two tunable impedance surfaces. (b)
The propagation properties for TM modes in an impedance waveguide
with two tunable impedance surfaces. The fundamental modes of metal
waveguide are plotted with dash-dot lines. $\beta_{\rm TE}$ and
$\beta_{\rm TM}$ refer here to the propagation constants of the TE
and TM modes, respectively. Only the two lowest modes are simulated}
\label{fig:dispersion2}
\end{figure}

The dispersion curve for a parallel-plate waveguide having two
tunable impedance surfaces is shown in Fig.~\ref{fig:dispersion2}.
Similar mode conversion is seen as in the case of just one impedance
surface. However, instead of a $180^\circ$ mode hop discussed
earlier, a $360^\circ$ mode hop occurs. This is simply because both
impedance surfaces induce a $180^\circ$ mode hop.

\begin{figure}[t!]
\centering \mbox{
\subfigure[]{\includegraphics[width=0.15\textwidth,height=5cm]{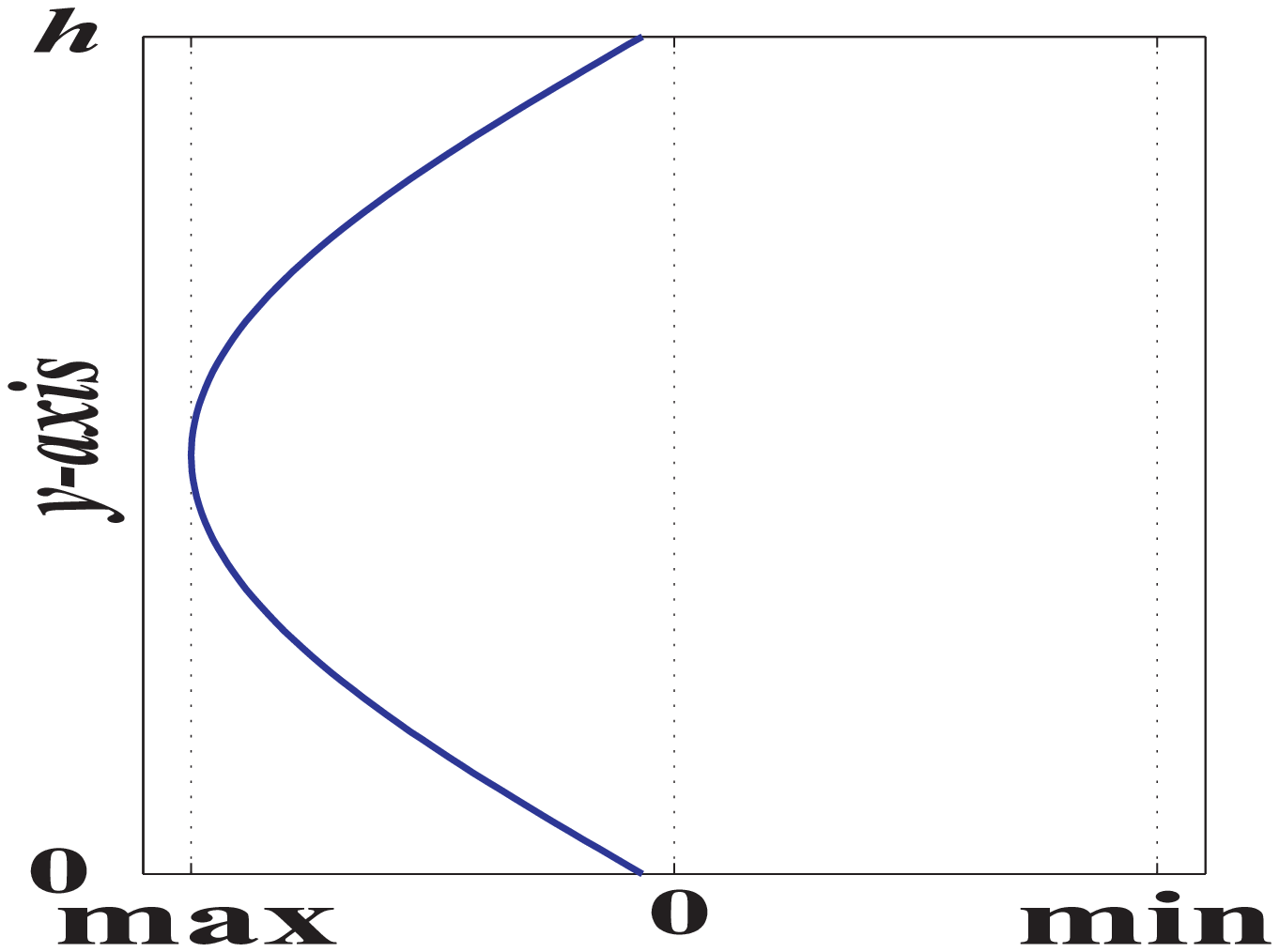}
}
\subfigure[]{\includegraphics[width=0.15\textwidth,height=5cm]{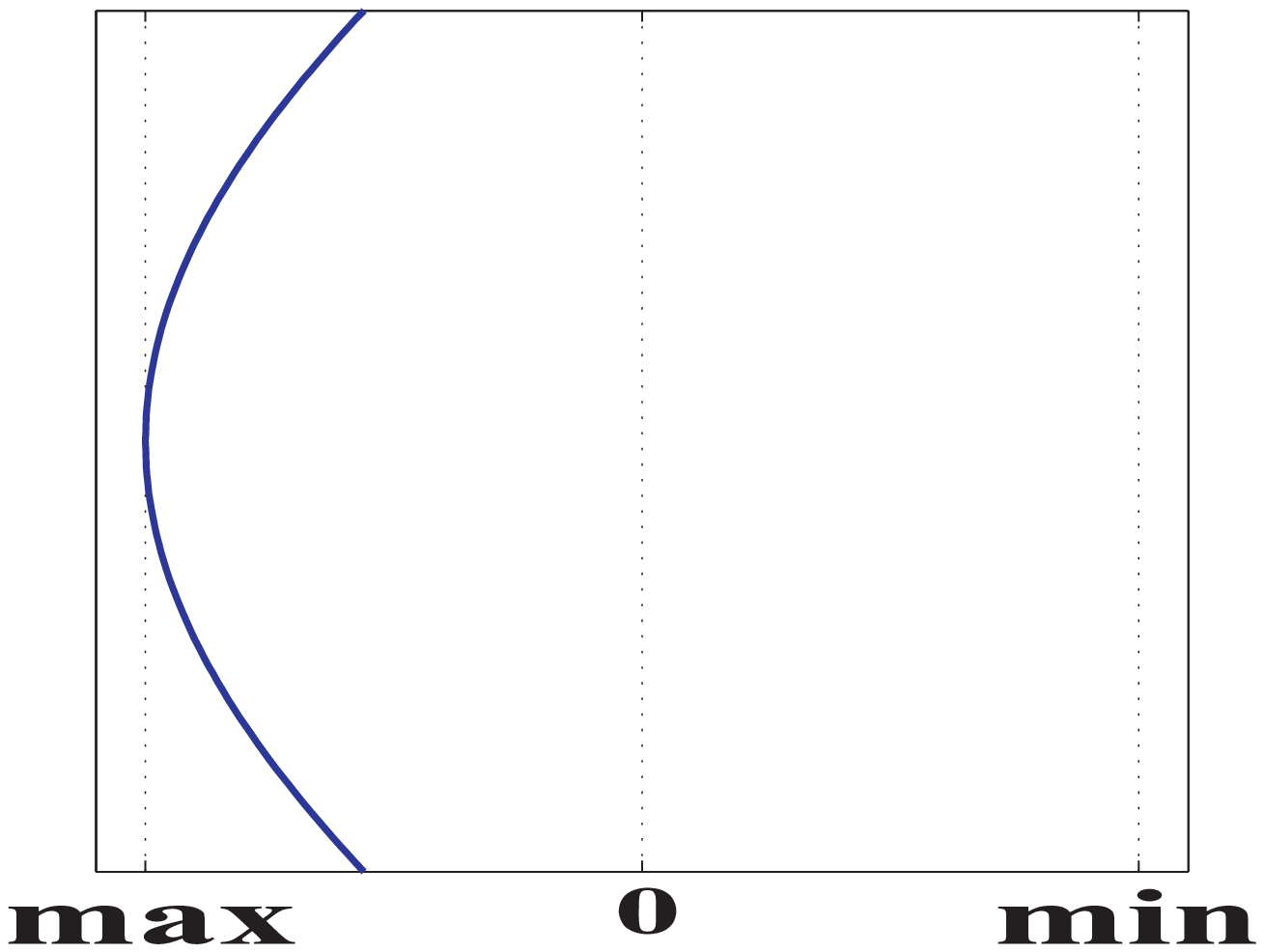}}
\subfigure[]{\includegraphics[width=0.15\textwidth,height=5cm]{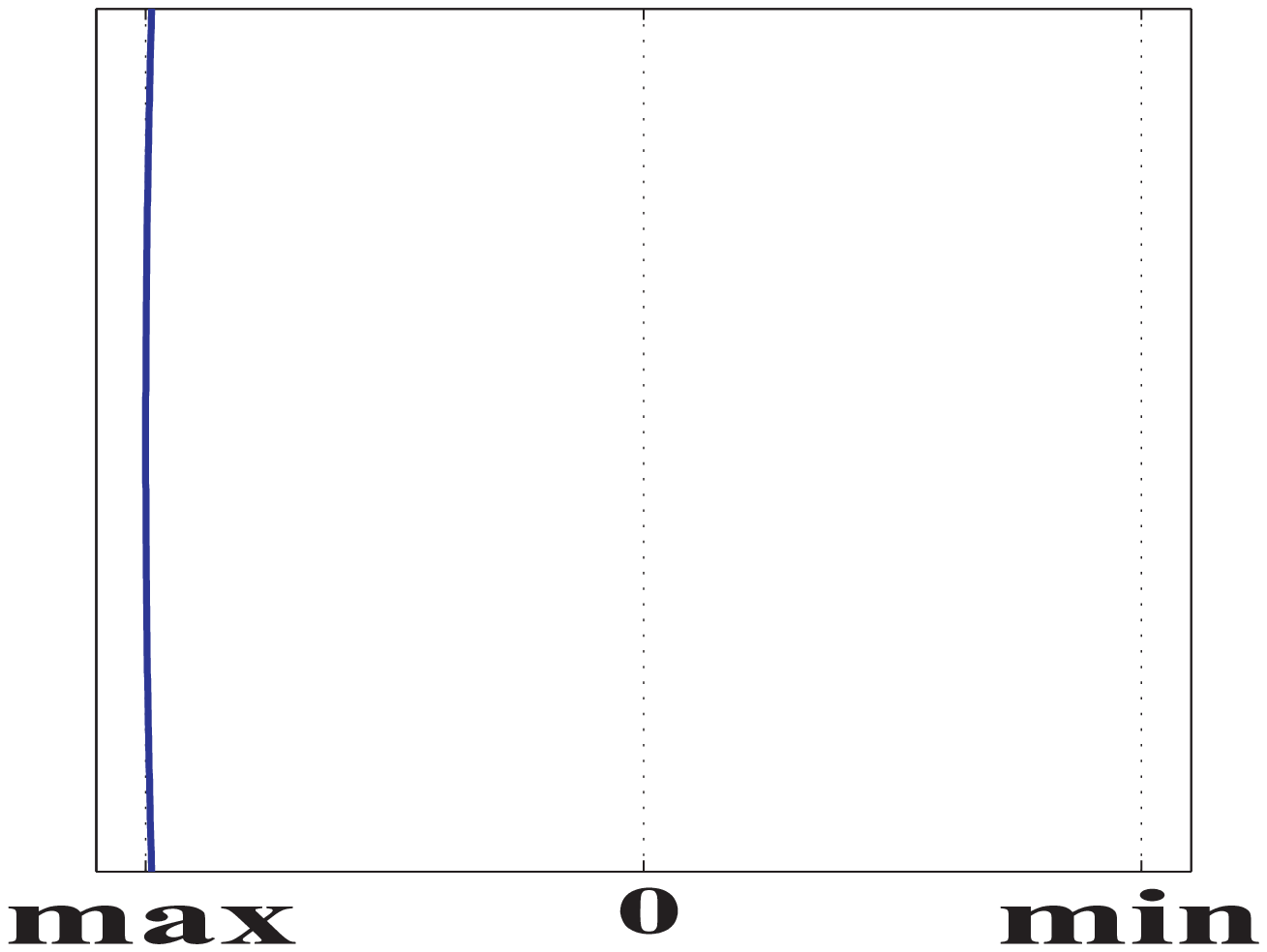}}}
\caption{The normalized magnitude of the $x-$component of the
electric field for the ${\rm TE}_{\rm 01}$ mode at points (a) $\beta
b=0.5$, $f = 24.9$\,GHz (b) $\beta b=0.69$, $f = 33.3$\,GHz, and (c)
$\beta b=0.71$, $f = 33.9$\,GHz  } \label{fig:fieldpatternTE01}
\end{figure}

\begin{figure}[t!]
\centering \mbox{
\subfigure[]{\includegraphics[width=0.15\textwidth,height=5cm]{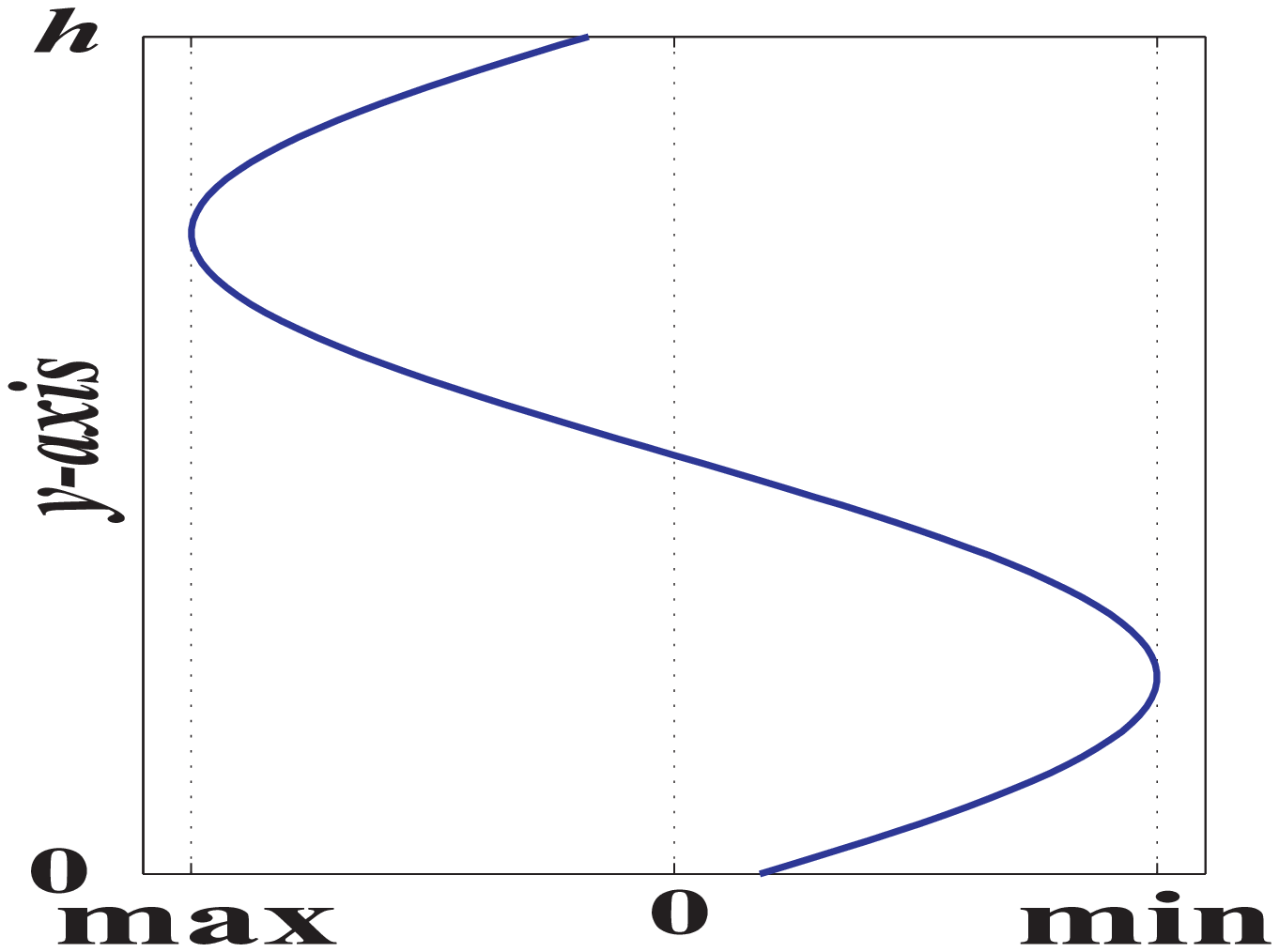}
}
\subfigure[]{\includegraphics[width=0.15\textwidth,height=5cm]{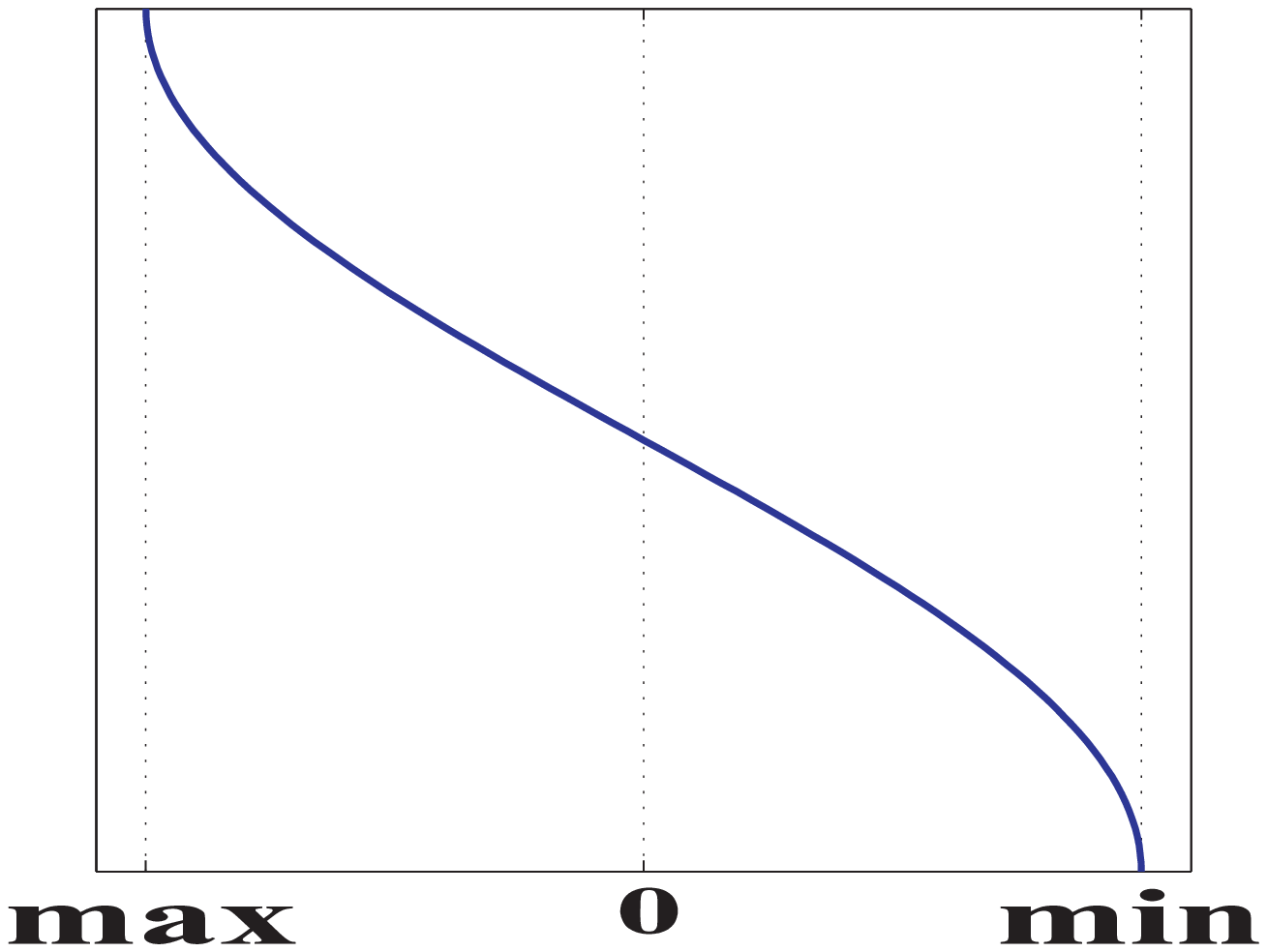}}
\subfigure[]{\includegraphics[width=0.15\textwidth,height=5cm]{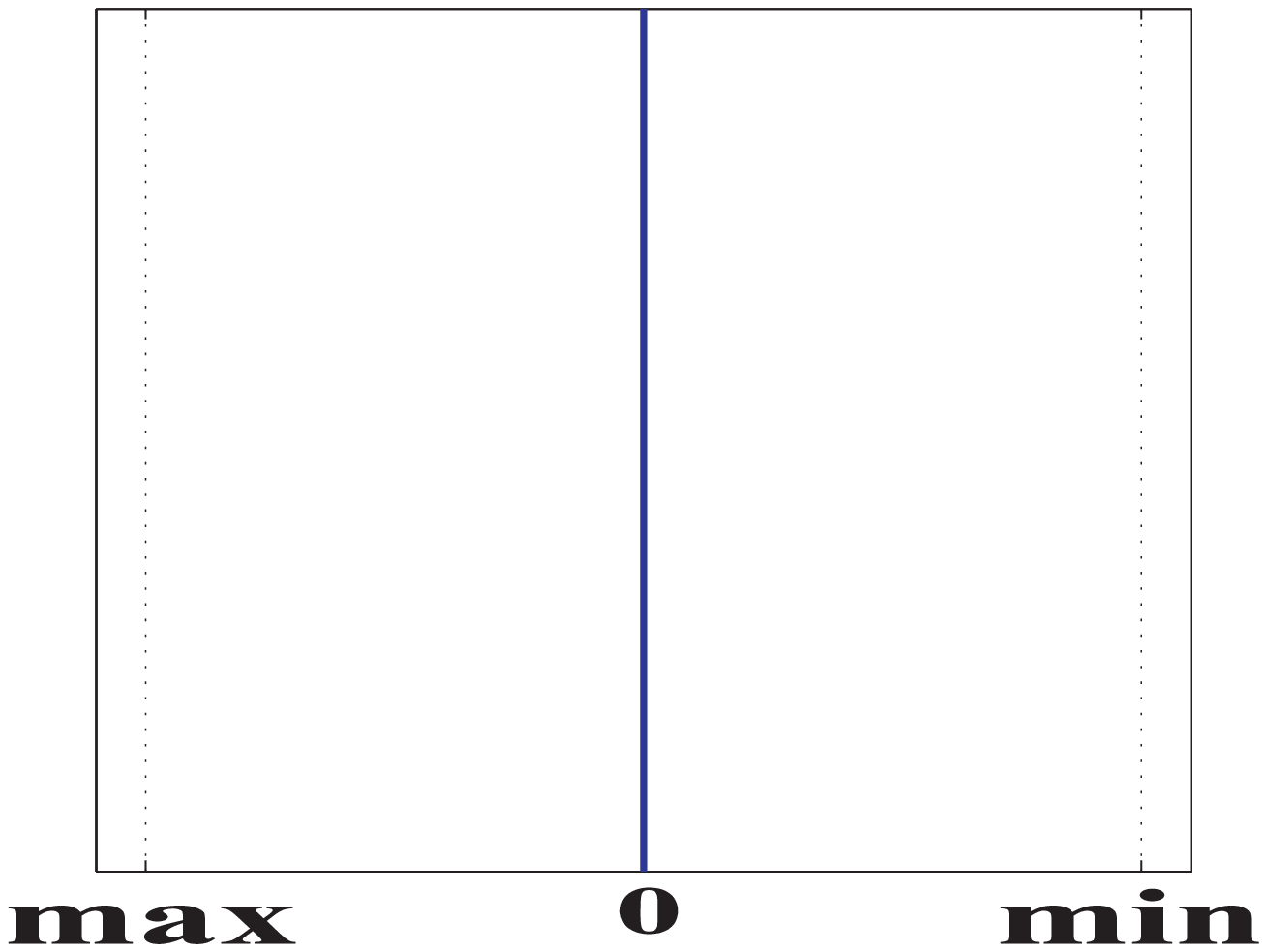}}}
\caption{The normalized magnitude of the $x-$component of the
electric field for the ${\rm TE}_{\rm 02}$ mode at points (a) $\beta
b=0.5$, $f = 27.7$\,GHz (b) $\beta b=0.69$, $f = 33.9$\,GHz, and (c)
$\beta b=0.77$, $f = 34.2$\,GHz  } \label{fig:fieldpatternTE02}
\end{figure}

In the case of two tunable impedance surfaces (see
Fig.~\ref{fig:dispersion2} (a)) interesting properties in the
vicinity of the resonance frequency of the surface are seen: Both
${\rm TE}_{\rm 01}$ and ${\rm TE}_{\rm 02}$ modes cross the light
line at only slightly different points. The ${\rm TE}_{\rm 01}$ mode
crosses the light line at $\beta b=0.71$, $f = 33.9$\,GHz whereas
the ${\rm TE}_{\rm 02}$ mode crosses it at $\beta b=0.77$, $f =
34.2$\,GHz. The field patterns of the ${\rm TE}_{\rm 01}$ and ${\rm
TE}_{\rm 02}$ modes are shown in Figs.~\ref{fig:fieldpatternTE01}
and \ref{fig:fieldpatternTE02}, respectively, for the case when
$C_{\rm var}=60$\,pF. Clearly, the modes can be divided into
symmetric and asymmetric modes which is not possible in the case of
just one tunable impedance wall.

\subsection{A single-mode waveguide}

\begin{figure}[t!]
\centering
\subfigure[]{\includegraphics[width=0.5\textwidth]{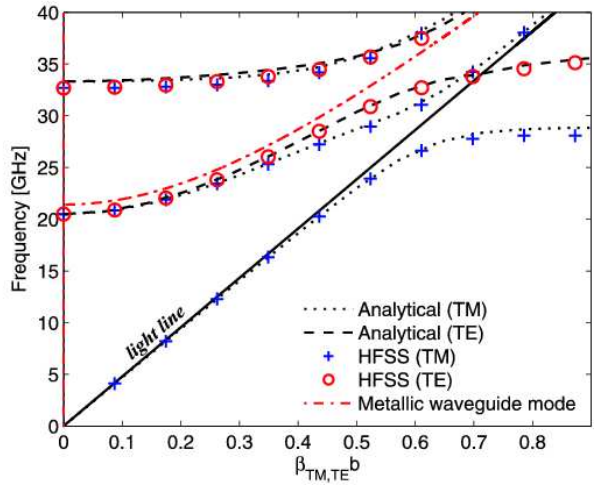}
}
\subfigure[]{\includegraphics[width=0.5\textwidth]{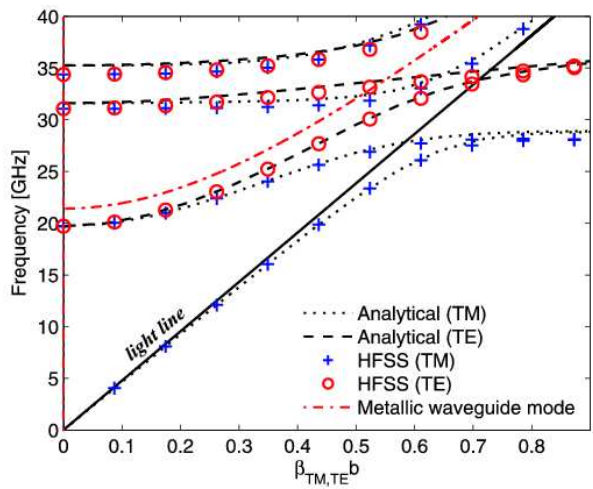}}
\caption{Color online. The propagation properties of an 7-mm high
impedance waveguide (a) with one tunable impedance surface and (b)
with two tunable impedance surfaces. The value of the varactor
capacitance is $C_{\rm var}=60$\,fF. The fundamental modes of metal
waveguide are plotted with dash-dot lines. $\beta_{\rm TE}$ and
$\beta_{\rm TM}$ refer here to the propagation constants of the TE
and TM modes, respectively.} \label{fig:dispersion_7mm}
\end{figure}

The dispersion curves of a 7-mm high waveguide having one tunable
impedance-surface are shown in Fig.~\ref{fig:dispersion_7mm} (a) and
in Fig.~\ref{fig:dispersion_7mm} (b) for a waveguide having two
tunable impedance-surfaces. In a regular metallic waveguide only one
mode would propagate in the waveguide in the Ka-band. However, for
the impedance waveguide, Fig.~\ref{fig:dispersion_7mm} clearly shows
two (three) modes in the case of one (two) high-impedance
sidewall(s). Because the response of the high-impedance surface for
the normal incidence is the same for both TE and TM modes, also the
cutoff frequencies are the same.

In Fig.~\ref{fig:dispersion_7mm} (a) the first TM mode diverges from
the first fundamental mode and converges to the light line as we
move through the resonance frequency of the high-impedance surface
up to higher frequencies. The first TE mode diverges also from the
first fundamental mode and crosses the light line at $f\approx34
$\,GHz (the resonance frequency of the high-impedance surface having
$C_{\rm var}=60$\,fF) after which the mode transforms into a
surface-wave mode. The second mode in the waveguide has the cut-off
frequency of $f=33.3$\,GHz. Both of these modes converge gradually
to the first fundamental mode.

In the case of two impedance surfaces (Fig.~\ref{fig:dispersion_7mm}
(b)) we can see two symmetric modes with cut-off frequencies of
$19.7$\,GHz and $35.3$\,GHz and one asymmetric mode with a cut-off
frequency of $31.6$\,GHz. Both the first symmetric and asymmetric TE
modes cross the light line at $f=33.8$\,GHz and $34.8$\,GHz,
respectively, and transform into surface-wave modes. The first
symmetric TM mode crosses the light line at $f=27.8$\,GHz and the
second symmetric TM mode converges to the first fundamental mode.
The asymmetric TM mode converges to the light line. This means that
there exists a stop band for both TM and TE modes. For TE modes the
width of the stop band is $1.5$\,GHz between the symmetric TE
waveguide modes and for the symmetric TM modes the width equals
$7.5$\,GHz.

\subsection{A below-cutoff waveguide}

\begin{figure}[t!]
\centering
\subfigure[]{\includegraphics[width=0.5\textwidth]{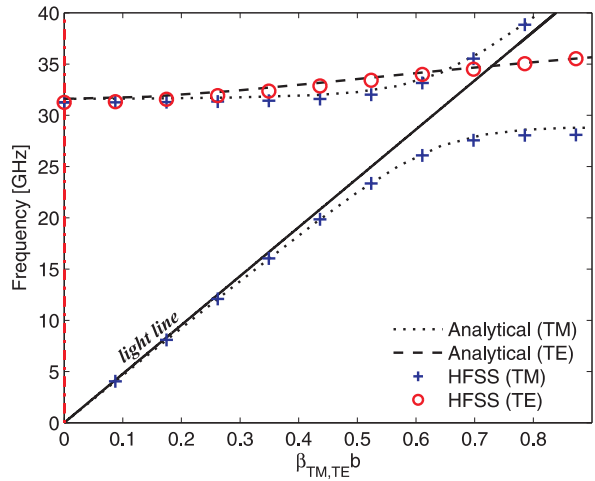}
}
\subfigure[]{\includegraphics[width=0.5\textwidth]{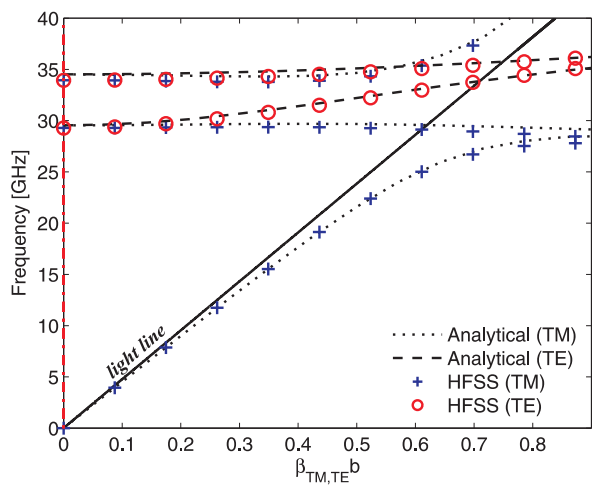}}
\caption{Color online. (a) The propagation properties of a
below-cutoff impedance waveguide with one tunable impedance surface.
(b) The propagation properties of a below-cutoff impedance waveguide
with two tunable impedance surfaces. The value of the varactor
capacitance is $C_{\rm var} = 60$\,fF. $\beta_{\rm TE}$ and
$\beta_{\rm TM}$ refer here to the propagation constants of the TE
and TM modes, respectively.} \label{fig:dispersion_35mm}
\end{figure}

In Figs.~\ref{fig:dispersion_35mm} (a) and (b) the dispersion curves
of a $3.5$\,mm-high parallel-plate waveguide are shown for the cases
of one and two impedance sidewalls, respectively.
Fig.~\ref{fig:dispersion_35mm} shows that although no TE mode would
propagate in the metallic waveguide below 42.8\,GHz, one mode
propagates in the impedance waveguide with one impedance sidewalls,
and two modes propagate in the case of two impedance sidewalls.
Similarly, in the case of TM modes, we find one or two extra modes
below the cut-off frequency depending on the number of impedance
sidewalls.

The TE modes form a narrow pass band both in the case of one and two
impedance sidewalls. In the case of just one impedance sidewall the
cutoff frequency of the first mode is $31.6$\,GHz and the TE mode
crosses the light line at $34.8$\,GHz. In the case of two impedance
sidewalls the cutoff frequencies for the symmetric and the
asymmetric modes equal $29.5$\,GHz and $34.5$\,GHz, respectively.
The symmetric mode crosses the light line at $33.9$\,GHZ and the
asymmetric mode crosses the light line at $35.8$\,GHz.

\begin{figure}[t!]
\centering
\includegraphics[width=0.5\textwidth]{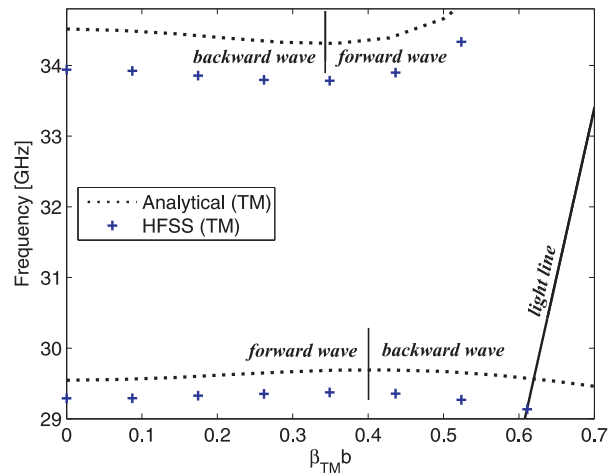}
\caption{A zoom-in of the Fig.~\ref{fig:dispersion_35mm}. The
backward wave propagation of the both symmetric and asymmetric TM
modes.} \label{fig:zoomin}
\end{figure}

In the case of two impedance sidewalls both propagating TM modes
demonstrate backward-wave propagation (see Fig.~\ref{fig:zoomin}).
The symmetric TM mode (cutoff frequency at $29.5$\,GHz) first
propagates as a forward wave but transforms into a backward wave
after the point $\beta d = 0.4$. The asymmetric mode (cutoff
frequency at $34.5$\,GHz) is first a backward wave and transforms to
a forward wave after the point $\beta d = 0.35$. In a 3.5\,-mm
metallic waveguide a TEM mode having the orientation of the field
components similar to the considered TM mode would propagate.
Figs.~\ref{fig:dispersion_35mm} (a) and (b) show that both with one
or two tunable impedance surfaces it is possible to create a tunable
stop band for the TEM mode.

\section{Discussion and conclusions}

An analytical model for a general type of tunable impedance surface
comprising an array of patches has been presented. Dispersion
equations for a parallel-plate waveguide having arbitrary surface
impedance sidewalls have been presented. Together with the presented
dispersion equations, the analytical model for the tunable
high-impedance surfaces is used to study the propagation properties
of impedance waveguide having either one or two tunable impedance
surfaces. In oversized waveguide mode conversion is shown. In
single-mode waveguide multi-mode propagation and band gaps are
shown. Furthermore, forward- as well as backward-wave propagation in
a below-cutoff waveguide is presented. The results are validated
with commercial full-wave simulation software. The concurrence
between the analytical and numerical results is very good.

The presented analytical model for the study of the propagation
properties in an impedance waveguide has proven to be very useful.
The model has been verified and used to predict the dispersion in
tunable impedance-wall waveguides in various example cases. The time
needed to produce the results with the analytical model compared to
the time consumed by numerical simulations is very marginal.

The dispersion results of the impedance waveguide show interesting
features. In oversized waveguides tunable impedance surfaces allow
one to realize tunable mode converters \cite{Thumm} and in many
other application, where field transforming inside waveguides is
needed. In addition, in single-mode or below the cutoff-frequency
waveguides possible applications for the tunable impedance surfaces
include phase shifters \cite{Higgins}, filters
\cite{Higgins_bandpass,Xin_bandstop}, and lenses \cite{Xin}. Without
doubt, the design work for all these applications benefit from an
efficient and simple model to predict the dispersion characteristics
of an impedance waveguide. This list can be continued with such
applications, as different types of antennas
\cite{Sievenpiper_beamsteering,Sievenpiper_leakywave,Hum}, tunable
artificial magnetic conductors (AMC) \cite{Feresidis2, Goussetis},
and tunable electromagnetic band-gap structures (EBG)
\cite{Feresidis,Goussetis, Yang}.

\end{document}